\documentclass[
reprint,
 aps,
 superscriptaddress,
bibnotes,
]{revtex4-2}

\usepackage{physics}
\usepackage{dsfont}
\usepackage{color,newfloat}
\usepackage{graphicx}
\usepackage{dcolumn}
\usepackage{bm}
\usepackage{amsmath,amssymb,amsthm}
\usepackage{mathtools}
\usepackage{multirow}
\usepackage{hhline}
\usepackage[dvipsnames]{xcolor}
\usepackage{hyperref}
\hypersetup{
    colorlinks=true,
    linkcolor=Maroon,
    citecolor=Maroon,
    urlcolor=Maroon
}
\usepackage[caption = false]{subfig}

\usepackage{makecell}

\begin{document}


\title{
High-threshold quantum computing by fusing one-dimensional cluster states
}

\author{Stefano Paesani}
\email{stefano.paesani@nbi.ku.dk}
\affiliation{Center for Hybrid Quantum Networks (Hy-Q), Niels Bohr Institute, University of Copenhagen, Blegdamsvej 17, Copenhagen 2100, Denmark}
\affiliation{NNF Quantum Computing Programme, Niels Bohr Institute, University of Copenhagen, Denmark.}

\author{Benjamin J. Brown}
\affiliation{IBM Quantum, T. J. Watson Research Center, Yorktown Heights, New York 10598, USA}
\affiliation{IBM Denmark, Pr\o vensvej 1, Br\o ndby 2605, Denmark}

\date{}


\begin{abstract}
We propose a measurement-based model for fault-tolerant quantum computation that can be realised with one-dimensional cluster states and fusion measurements only; basic resources that are readily available with scalable photonic hardware.
Our simulations demonstrate high thresholds compared with other measurement-based models realized with basic entangled resources and two-qubit fusion measurements.
Its high tolerance to noise indicates that our practical construction offers a promising route to scalable quantum computing with quantum emitters and linear-optical elements.
\end{abstract}

\date{\today}

\maketitle


\noindent\textbf{Introduction.}
Scalable quantum-computing architectures~\cite{Dennis02, raussendorf2007} are built on quantum error-correcting codes~\cite{Shor96} that identify and correct for errors that quantum hardware experiences as logical algorithms progress. 
However, it remains difficult to produce an architecture with a sufficiently large number of high-quality qubits to complete long quantum computations reliably.
To overcome this technological challenge, we should design bespoke quantum architectures that take advantage of the positive features of scalable physical hardware~\cite{Fowler2012, Nickerson2014, Nemoto2014, GimenoSegovia2015, Lekitsch2017, Herr2018, Chamberland2020, Darmawan2021practical, Bartolucci2021fusion, Bourassa2021blueprintscalable, Larsen2021, Wan2021fault, Chamberland2022, Claes2022, Strikis2022}. 
Ideally, we should exploit the native operations of the underlying quantum system to minimize the noise processes that will lead to computational errors.

Quantum emitters are a promising tool to implement photonic architectures~\cite{uppu2021quantum, lu2021quantum}.
They enable the deterministic generation of a variety of entangled states with even a single emitter~\cite{Lindner2009proposal, Li2022}, such as one-dimensional cluster states~\cite{Raussendorf2001, Lindner2009proposal}, via appropriate pulse sequences driving a spin-photon interface interleaved with photon emission.
Quantum emitters have been demonstrated with several hardware platforms including quantum dots~\cite{Schwartz2016, Istrati2020}, atomic systems~\cite{Yang2022, Thomas2022},  superconducting circuits~\cite{besse2020realizing}, and nitrogen-vacancy centers~\cite{vasconcelos2020}. Indeed, deterministic entanglement has recently been demonstrated between as many as 14 photons~\cite{Thomas2022}. 

Measurement-based models for fault-tolerant quantum computing~\cite{Raussendorf2006} are particularly well suited for photonic architectures. In this model we prepare entangled resources that can be produced by constant-depth circuits.
Measuring these resources process quantum information and, at the same time, extracts syndrome data for error correction. 
We often consider realizing the resource states by entangling individual physical qubits~\cite{Nemoto2014, Herr2018} or small, constant-sized, entangled resources~\cite{GimenoSegovia2015, Bartolucci2021fusion}. 
These proposals place a significant demand on optical hardware to entangle the individual physical systems, using either unitary gates or fusion-based operations~\cite{Browne2005resource}.

\begin{figure}
    \centering
    \includegraphics{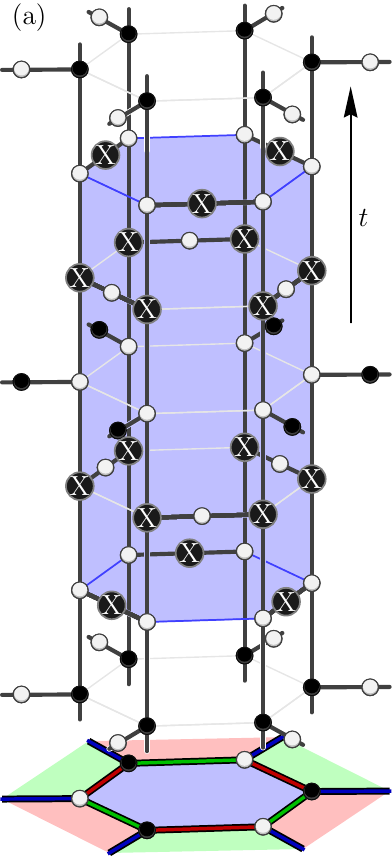} \includegraphics{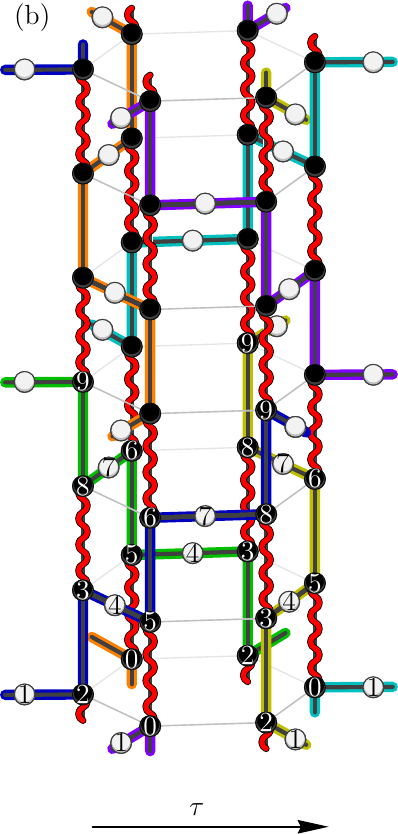}
   \includegraphics{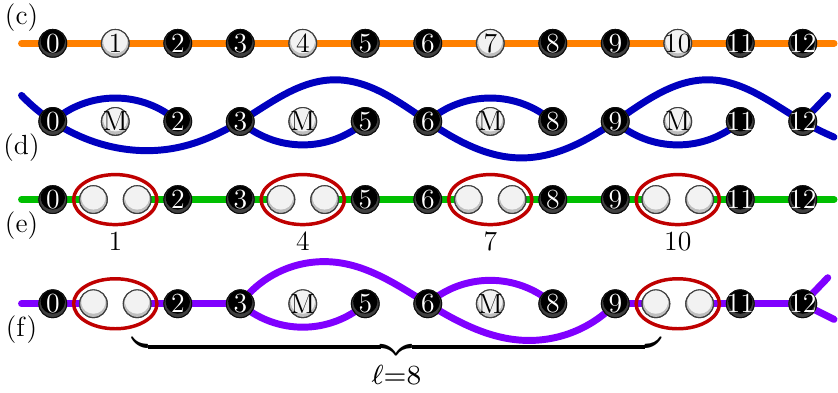} 

    \caption{Realizing the foliated Floquet color code with linear cluster states and fusion measurements. (a)~The resource state $\mathcal{R}$ we construct is illustrated by graph. Detectors $S \in \mathcal{S} = \mathcal{R} \cap \mathcal{M}$ are the product of Pauli-X terms at the boundaries of local cells. The vertical time-like axis is labeled $t$ that indexes layers of the three-dimensional structure.
    (b)~The resource state can be composed of one-dimensional cluster states, or `chains', as in~(c), shown by thick solid lines, and fusion measurements, marked by wavy red lines. The qubits of each chain are indexed with label $\tau$.
    We can make variations of our construction with different input resources. We identify the qubits of the chain in (c) with those in (d-f) with their numerical indices.     
    We obtain branched chains~(d), up to local Clifford operations, by measuring white qubits of (c) in the Pauli-X basis. We can also produce long chains from small resource states by fusing the first and last qubit of linear chains of length $\ell=4$ for example, (e). We can also fuse the end points of short branched chains, (f), where we show a short branch of $\ell=8$ qubits.
    \label{fig:stabilizers}
    }
\end{figure}

Here, we find that we can reduce the complexity of realizing entangled resources by taking advantage of the one-dimensional entangled states we can produce readily with quantum emitters. The remaining entangling operations that are needed to perform measurement-based fault-tolerant quantum computing are made at the readout stage using two-qubit fusion measurements~\cite{Browne2005resource}. Central to our architecture is a specific resource state whose geometry is obtained by foliating~\cite{Bolt2016foliated, Brown2020universal} the Floquet color code~\cite{Davydova2022, Kesselring2022}; an example of a dynamically driven code~\cite{Hastings2021dynamically}. These codes are of recent interest due to their high thresholds~\cite{Gidney_2022, Paetznick2023} and their implementation with weight-two parity measurements.
Our numerical simulations demonstrate very high thresholds by refining our architecture. Specifically, we propose using a decorated or `branched' one-dimensional cluster states to reduce the number of qubits we need to prepare and measure in the fault-tolerant system. In what follows we describe our construction before presenting numerical results.

\ \\
\noindent\textbf{The foliated Floquet color code.}
In measurement-based fault-tolerant quantum computation we prepare a resource state that can be specified by a graph~\cite{Raussendorf2001, hein2006}. A qubit is initialised in an eigenstate of the Pauli-X matrix on every node of the graph $v$. Controlled-phase gates are then applied to pairs of qubits whose corresponding nodes share an edge. The stabilizer group~\cite{Gottesman97} of the resource state $\mathcal{R}$ is generated by elements $R_v = X_v \prod_{q \in \Delta v} Z_q $ for each node $v$, with $\Delta v$ the neighbourhood of $v$ and $X_v,\, Z_v$ the standard Pauli matrices acting on $v$. The resource is then measured, projecting it onto the state with stabilizer group $\mathcal{M}$, generated by the commuting Pauli measurements we make. The measurements provide syndrome data $\mathcal{S} = \mathcal{R} \cap \mathcal{M}$ that is used to identify errors~\cite{Raussendorf2005, Brown2020universal, Bell2022}. We find our measurement-based model by foliating~\cite{Bolt2016foliated, Brown2020universal} the Floquet color code~\cite{Davydova2022, Kesselring2022}. The resulting graph is shown in Fig.~\ref{fig:stabilizers}(a). We call our model the foliated Floquet color code (FFCC).

Foliation is a method for mapping circuit-based models, that are realized with static qubits, onto the measurement-based picture. Roughly speaking, data qubits of the static model are replaced with linear cluster states, where the ordering of the qubits of these linear clusters can be regarded as a time-like direction in the static picture. Measurements that are made in the static picture are realized in the measurement-based picture by coupling additional `check qubits' to the appropriate qubits of the linear clusters that model the time evolution of data qubits of the static model.

We obtain a three-dimensional lattice by applying the foliation methods described in Ref.~\cite{Brown2020universal} to the Floquet color code~\cite{Davydova2022, Kesselring2022}, where data qubits of lie on vertices of a two-dimensional hexagonal lattice, shown at the base of Fig.~\ref{fig:stabilizers}(a). The graph describing $\mathcal{R}$ has two types of nodes; let us call them data nodes and check nodes. Data nodes have indices $(q,t)$ where $1 \leq q \leq n$ index data qubits of the hexagonal lattice. The second index $ 1 \leq t \leq T$ denotes a temporal order. Data nodes $(q,t)$ and $(q,t+1)$ share an edge for all $q$ and $t$.

\begin{figure}
    \centering
    \includegraphics{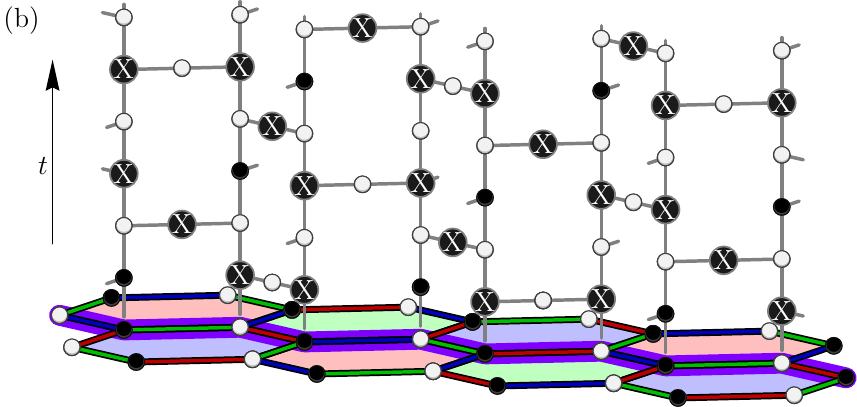}
    
    \includegraphics{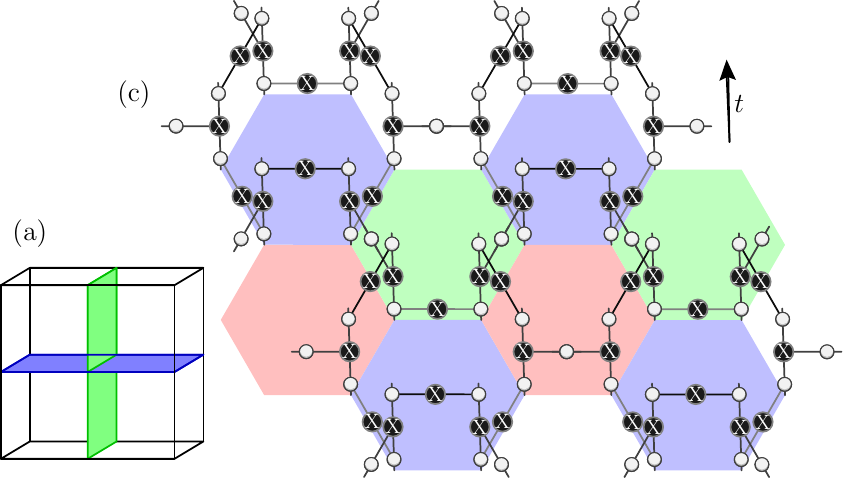}
    \caption{Correlation surfaces in the foliated Floquet color code. (a)~Shows the orientation correlation surfaces that lie orthogonal to a canonical spatial direction and the temporal direction in green and blue, respectively. We show the microscopic details of these operators in (b) and (c), respectively, with time-like axes, $t$, marked with arrows.}
    \label{fig:CorrelationSurfaces}
\end{figure}

The remaining edges to complete the graph for $\mathcal{R}$ connect data nodes to check nodes. The check nodes are associated with the edges of the hexagonal lattice.  There are three types of edges, that are assigned colors red, green or blue~\cite{Bombin2006}. Specifically, we three-color the faces of the hexagonal lattice such that no two adjacent faces have the same color. Edges of the hexagonal lattice are then assigned the color of the two faces they connect. Let us also denote $\{v,w\} = \partial e $ the pair of vertices of the hexagonal lattice that are connected by edge $e$. We have check nodes associated to edges of different colors at different times. Let us denote the check nodes with indices $(e,t)$. We have check nodes associated only to the blue, green and red edges at times $3t+1$, $3t+2$ and $3t+3$, respectively. Every check node $(e,t)$ of the appropriate color then shares and edge with the two data nodes $(v,t)$ for both $v \in \partial e$. These check nodes are entangled to the data nodes to correspond precisely to the measurement sequence of the Floquet color code according to the foliation methods of Ref.~\cite{Brown2020universal}. This completes the construction shown in Fig.~\ref{fig:stabilizers}(a).

Up to lattice geometry, the model we have produced shares many features with the topological cluster-state model~\cite{Raussendorf2005} where, for now, we assume we measure all of the qubits in the Pauli-X basis, i.e., we project the system onto the stabilizer group $\mathcal{M} = \langle \pm X_v \rangle$ to obtain the code detectors that identify errors.  The FFCC has local detectors $\mathcal{S} = \mathcal{R} \cap \mathcal{M}$ on cells of the lattice. We show an example of a local detector in Fig.~\ref{fig:stabilizers}(a). Similiarly, the foliated code gives rise to correlation surfaces that propagate quantum information between input and output regions~\cite{Raussendorf2003, Raussendorf2005, raussendorf2007}. We show examples of correlation surfaces in Fig.~\ref{fig:CorrelationSurfaces}. Additionally, our model can be divided into two disjoint lattices of qubits; the primal and dual lattice, where detectors and correlation surfaces are supported on only one of the two disjoint lattices. The primal and dual lattices are distinguished with black and white vertices in Fig.~\ref{fig:stabilizers}(a) and Figs.~\ref{fig:CorrelationSurfaces}(b) and~(c).

The common features of our model with that in Ref.~\cite{Raussendorf2005} mean that we can adopt the fault-tolerant gate set presented in Ref.~\cite{raussendorf2007} by reconfiguring our measurement pattern such that gates are performed by braiding different types of defect punctures, and by distilling magic states to complete a universal gate set. It may be interesting to consider adapting the methods of Ref.~\cite{Herr_2018lattice, Herr2018, Brown2020universal, Bombin2021logical} to our lattice geometry for more general gate operations based on the braiding of twist~\cite{Bombin2010topological} and corner~\cite{Brown2017poking, Brown2020universal} defects. It may also be interesting to investigate implementations of non-Clifford gates with this lattice geometry~\cite{Bombin2018, Brown2020a}.

\ \\
\noindent\textbf{Quantum computing with fusion measurements.}
Let us now look for practical ways of preparing and measuring the resource state $\mathcal{R}$. Recently Ref.~\cite{Bartolucci2021fusion} has shown that we can eliminate the difficulty of preparing a large entangled resource state by completing the preparation and readout of a resource state with probabilistic entangling Bell measurements, i.e. fusion operations~\cite{Browne2005resource}. With an appropriate choice of fusion measurements, we find that we can decompose the graph shown in Fig~\ref{fig:stabilizers}(a) into a series of physical one-dimensional cluster states, that we call chains, and fusion measurements only, see Fig.~\ref{fig:stabilizers}(b). The qubits of the chain are indexed $ 0 \le \tau \le 3 T-1$, see Fig.~\ref{fig:stabilizers}(c) where a single chain is laid out independent of the three-dimensional construction. In the figure each chain is marked by bold lines of different colors where we see that every chain has three qubits at each time step $t$.

It is helpful to bicolor the vertices of the hexagonal lattice black and white such that no two vertices of the same color share an edge. Likewise, data nodes $(v,t)$ have the same color as vertex $v$ of the hexagonal lattice. We identify all of the black qubits $(v,1)$ of the foliated system at $t=1$ with the $\tau = 0$ qubit of each chain. The next qubit of the chain with $\tau = 1$ is identified with the unique edge qubit $(e,1)$ with $ v\in \partial e$ and then qubit $\tau = 2$ is identified with $(w,1)\not= (v,1)$ with $w \in \partial e$. The chain then progresses to the next level before repeating, where the $\tau = 3$ element of the chain is identified with $(w,2)$. The progression continues {\it ad infinitum}. In general we have that the  $ \tau = 3t$($3t+1$)-th qubit lies at qubit $ (v, t)$ ($ (e,t) $ with $v \in \partial e$), and the $3t + 2$-th qubit of the chain lies at $(w,t)$ and $w \in \partial e$ with $w \not= v $. The next qubit in each chain lies at qubit $(w,t+1)$. Indices $ (v,t)$ and $(w,t)$ are, respectively, black and white (white and black) for odd (even) values of $t$.

By comparing Figs.~\ref{fig:stabilizers}(a) and~(b) we see the chains are organized such that many of the edges of the resource state graph are completed in the production of the chains. However, some entangling operations remain to be performed. We then complete the resource state with fusion operations.
Up to local Clifford operations, we can interpret a successful fusion measurement as a controlled-phase gate, i.e. creation of an edge, followed by two single-qubit measurements in the Pauli-X basis.
Fusion measurements shown in Fig.~\ref{fig:stabilizers}(b) therefore complete the entangling operations needed to produce the resource state and subsequently make the single-qubit measurements we need for readout. Specifically, we perform fusion measurements between black (white) data nodes $(v,t)$ and $(v,t+1)$ at odd (even) values of $t$.
Final read out in single-qubit bases different from Pauli-X, required for universal gate sets, can also be simply implemented by reconfiguring the linear optical fusion circuit used~\cite{GimenoSegovia2015, Bartolucci2021fusion}.

We consider variations where chains are replaced by other resources that may be more readily implemented.
In Fig.~\ref{fig:stabilizers}(c) we show a single chain, where the data(check) nodes are marked black(white). As Fig.~\ref{fig:stabilizers}(b) shows, the check nodes are completely entangled to their neighbours $\partial e$ when the chain is produced. 
We may therefore consider replacing the chain with a decorated chain that is the post-measurement state that would be obtained if the check nodes of a standard chain are measured in the Pauli-X basis.
Up to local Clifford operations~\cite{hein2006} we obtain the resource state with branches, as shown in Fig.~\ref{fig:stabilizers}(d). This state is readily prepared with quantum emitters, see Appendix~\ref{app:branched_chains}.
We call the measurement-based model realized with these decorated chains the {\it branched construction}.
The branched construction therefore neglects to produce check nodes and only instead construct the data nodes, such that the edge measurements are effectively already made.

We might also look for variations where we produce the large resource states from smaller entangled resources. For example, fusion measurements between qubits of $\ell = 4$-qubit  linear cluster states, as shown Fig.~\ref{fig:stabilizers}(e), produce the chains we use in our construction. We can also interpolate between the short chain construction and the branched chain construction by connecting finite-size branched chains at their endpoints via fusion measurements. We show a short branched chain of length $\ell = 8$ qubits in Fig.~\ref{fig:stabilizers}(f).

\ \\
\noindent\textbf{Error correction.}
We can also adopt methods used for the topological cluster state to perform error-correction with the FFCC. We are interested in correcting Pauli errors as well as heralded qubit erasure occurring on the physical qubits of the system.

Every single qubit supports exactly two detectors~\cite{Kesselring2022}. It follows that, if a Pauli error occurs, two detectors are violated. We can therefore regard a single error as a string-segment with violated detectors at its end points~\cite{Dennis02, Raussendorf2005}. In general, multiple Pauli errors compound to make multiple strings of potentially greater length. We can correct the errors by finding pairs of violated detectors that are corrected by short string operators. Provided the proposed correction has an equal parity of errors supported on the correlation surface as the initial error, we declare the correction successful. We find nearby pairs of violated detectors using \texttt{PyMatching}~\cite{higgott2022}; an implementation of minimum-weight perfect matching~\cite{Dennis02}.

We also adopt the methods presented in Ref.~\cite{Barrett2010} for dealing with heralded erasure. If a qubit $v$ is erased, we no longer have access to its two stabilizers $S_{b}$ and $S_c$ that support $v$. To deal with this, we neglect the erased qubit, and we replace these two supporting stabilizers with their product $S_bS_c$, thereby creating a super cell. In general we have to update the lattice to produce super cells for all the qubits that experience erasure. Error correction for Pauli errors on the updated lattice then proceeds in the same way using super cells on qubits that have not been erased. It is also important to find a correlation surface that contains no erased qubits. To do so, we multiply the correlation surface by detector operators to find a variation of the operator such that no qubits are erased, which can be readily achieved by Gaussian elimination~\cite{Connolly2022}. Otherwise, we consider error correction to fail.

The error-correction methods described above are readily adapted for the fusion error model~\cite{Bartolucci2021fusion}.
In this model, a fusion erasure takes into account the erasure of measured qubits as well as the possible failure of the fusion measurement. The fusion error model also includes fusion measurement errors induced by Pauli errors on the fused qubits.

 \begin{figure}
  \centering
  \includegraphics[
  width=0.48 \textwidth]{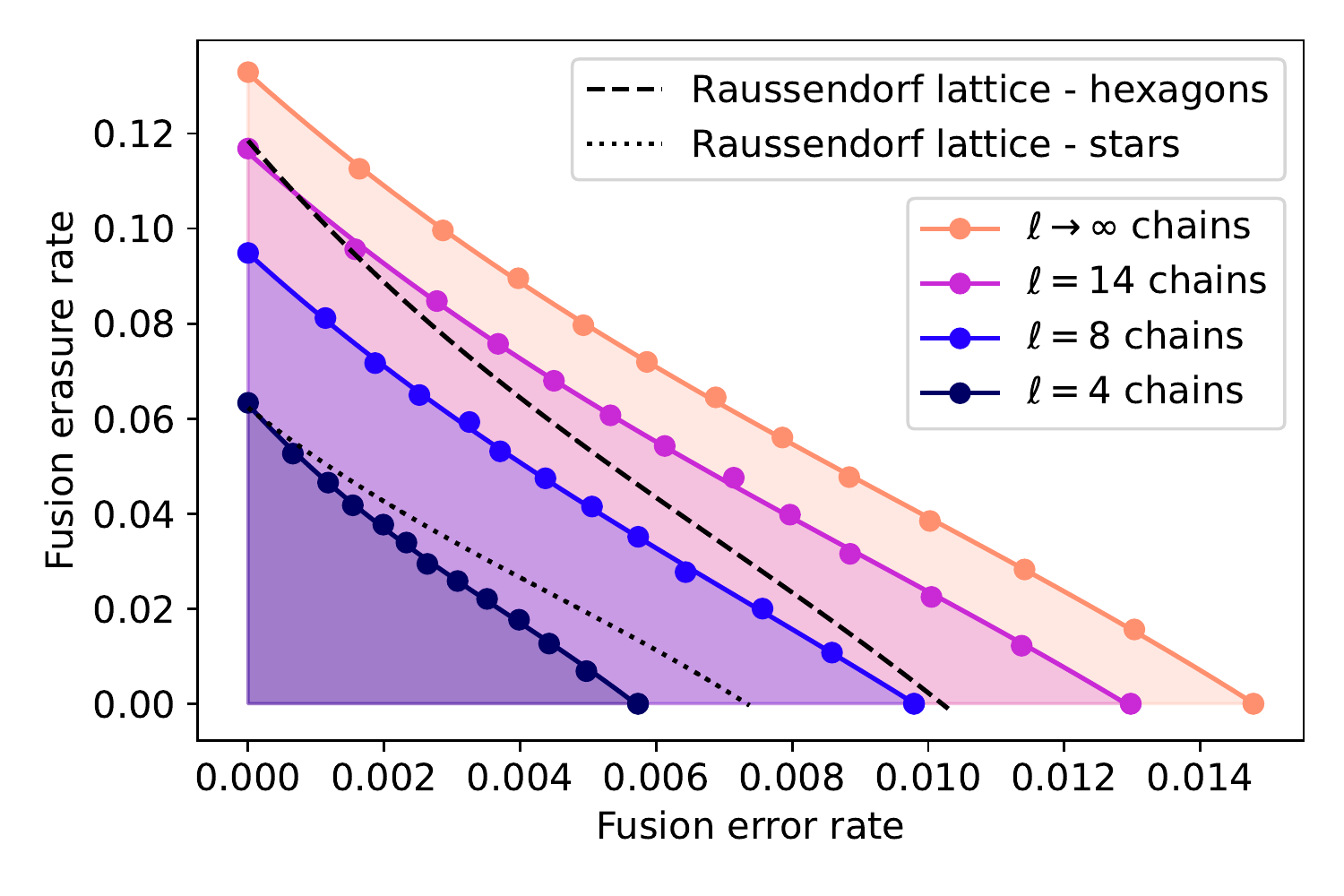}
  \caption{
Fault-tolerant regions for fusion-based constructions. 
Solid lines show thresholds when constructing the FFCC lattice by fusing branched chains with length $\ell \in \{4, 8, 14\}$ and for the limit $\ell \rightarrow \infty$ where the length is much longer than the lattice unit cell size.
For comparison, we also show the performance of the constructions from Ref.~\cite{Bartolucci2021fusion} using hexagonal and four-qubit star-shaped resource states by the dashed and dotted black lines, respectively.
}
  \label{fig:thresholds_fusion}
\end{figure}

\ \\
\noindent\textbf{Threshold estimates.}
We evaluate threshold error rates for a phenomenological  fusion-based noise model as in Ref.~\cite{Bartolucci2021fusion} where erasures and errors occur independently for each fusion measurement and with a probability equal to the associated noise rate.
Fusion networks are simulated with periodic boundaries so that we can check logical failures for the two distinct correlation surfaces shown on orthogonal planes in Fig.~\ref{fig:CorrelationSurfaces}.
Thresholds are evaluated by comparing the logical error rate of our decoder for different
noise parameters and different lattice sizes. We evaluate logical failure rates with $10^4$ Monte-Carlo samples (see Appendix~\ref{app:decoding} for details).
We also report on an analysis for bare lattices in Appendix~\ref{app:lattice_thresholds}.
We show threshold error rates for different constructions in Fig.~\ref{fig:thresholds_fusion} for various rates of error and erasure that occur under fusion measurements.
The highest thresholds are obtained using the branched chain construction. The threhsold interpolates between an erasure rate $\sim 13.2\%$ to an error rate $\sim1.5\%$.
The thresholds we obtain outperform previous constructions based on hexagonal and star-shaped resource states used to produce Raussendorf lattice structures~\cite{Bartolucci2021fusion}. We reproduced the thresholds for these models. They are shown by dashed and dotted lines in Fig.~\ref{fig:thresholds_fusion}.
Such improvements manifest the advantages of constructing a model with a lower valency lattice, a direct consequence of having only weight-two parity measurements,  while still inheriting good error-correction performances of the underlying Floquet color code.     
The thresholds for constructions that include fusing branched chains of constant size, as shown in Fig.~\href{fig:stabilizers}{\ref{fig:stabilizers}(e-f)}, are also reported in Fig.~\ref{fig:thresholds_fusion} for different lengths $\ell \in \{4, 8, 14\}$.
Significant improvements over previous constructions can be observed already for chains with a moderate constant size of 14 qubits.
Thresholds relative to erasure can also be significantly improved by biasing the fusion failures, as recently shown in Refs.~\cite{sahay2022tailoring,bombin2023increasing}. In Appendix~\ref{app:biased} we report how these techniques can enhance erasure thresholds also in our constructions.

\ \\
\noindent\textbf{Discussion.}
To summarize, we have demonstrated a practical architecture to realize fault-tolerant measurement-based quantum computation using one-dimensional entangled resources.
 Such resource states are readily and deterministically prepared with quantum emitters, avoiding the large overheads required for preparing entangled resource states via multiplexed probabilistic processes in all-optical approaches~\cite{bartolucci2021switch} and can also be relevant for other physical systems with probabilistic entangling operations~\cite{Monroe2014, Roch2014, ruf2021}. 
Furthermore, our numerical simulations demonstrate very high thresholds. We obtained these results focusing on phenomenological noise that models all noise sources with a single parameter. This allows us to compare our proposal with others already presented in the literature. In the future, it will be important to run simulations that consider more representative models of noise sources that we anticipate in the laboratory. In Appendix~\ref{app:physical_noise} we discuss some of them in the context of preparing linear cluster states with quantum emitters.

We argue that the high-thresholds we have obtained are due to the resource efficient and practical construction we have designed that requires a relatively low number of fusion operations for its realization. 
Better models with higher noise tolerance could, perhaps, be obtained by finding more general lattice structures, see e.g. Refs.~\cite{nickerson2018, Newman2020generatingfault}. Ultimately, we may find more robust models by finding more general resource states that can be readily produced, for example, with interactions between multiple emitters~\cite{gimeno2019, Li2022}.


\ \\
\textit{Acknoweldgements.}
We are grateful for conversations with M. L\"obl, L. Pettersson, A. S\o rensen and Y. Zhang. 
S.P. acknowledges funding from the Cisco University Research Program Fund (nr. 2021-234494) and the Marie Skłodowska-Curie Fellowship project QSun (nr. 101063763). B.J.B. received support from the U.S. Department of Energy, Office of Science, National Quantum Information Science Research Centers, Co-design Center for Quantum Advantage (C2QA) under contract number DE-SC0012704.  B.J.B. is grateful for the hospitality of the Center for Quantum Devices at the University of Copenhagen.

\bibliography{biblio.bib}


\newpage 
\clearpage

\pagenumbering{arabic}

\appendix

\renewcommand{\thesection}
{\Alph{section}}

\renewcommand{\thefigure}{A\arabic{figure}}
\setcounter{figure}{0} 

\renewcommand{\thetable}{A\arabic{table}}
\setcounter{table}{0}


\section{Generation of branched chains via a single quantum emitter}
\label{app:branched_chains}

\begin{figure}[b]
  \centering
  \includegraphics[
  width=0.48 \textwidth]{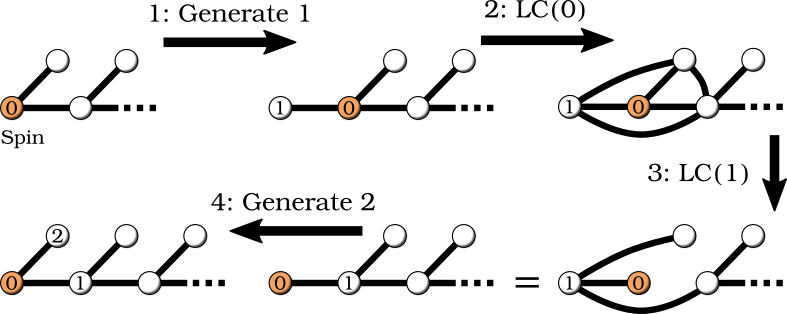}
  \caption{
Steps in the recursive procedure for the generation of branched chains with a quantum emitter. 
Each recursion adds a branched layer in the one-dimensional cluster.
In each recursion, we (step 1) generate a new photon with the quantum emitter, which effectively adds a new leaf to the graph state attached to the spin node.  
Then (step 2) local operations on the spin and photons are performed to perform a local complementation (LC) operations on qubit 0 and (step 3) on qubit 1, which correspond to the graph transformations shown.
Finally, (step 4) by generating a new photon we again have a branched chain with now an additional layer. 
Repeating these steps generates a branched chain of arbitrary length.
}
  \label{fig:branched_chains}
\end{figure}

%
Here we describe how to generate branched chain clusters with a single quantum emitter. We consider the recursive procedure as described in Fig.~\ref{fig:branched_chains} and its caption.
The procedure describes the generation in the graph state picture, where the spin-controlled photon generation at the quantum emitter corresponds to adding a leaf, i.e. a single-edged node, to the node associated to the spin qubit. The qubit associated to the spin is shown in orange in the figure.
We also use a graph transformation called local-complementation (LC) operation. The operation, that acts on a node of the graph, applies the complementary subgraph of its neighboring qubits.
Notably, local complementation can be implemented via local Clifford gates~\cite{hein2006}. It, therefore, corresponds to single-qubit operations on the spin qubit and on the photons, which can be readily implemented deterministically.
The procedure can thus be readily translated into a sequence of photon generation operations by the quantum emitter, that are interleaved with single-qubit gates on the spin, as well as local operations on the emitted photons. 
Finally, we note that, while the procedure generates branches with a single leaf per branch, it can be easily expanded to multiple leaves per branch.
It is in fact sufficient to repeat the photon generation multiple times before doing the local complementation steps.
Such states, which can be seen as a more hybrid version in between star graphs and linear chains, present nodes with higher valency, and could thus be investigated for the resource-efficient construction of more general lattices.

\section{Threshold estimates for lattices}
\label{app:lattice_thresholds}

 \begin{figure}[b]
  \centering
  \includegraphics[
  width=0.48 \textwidth]{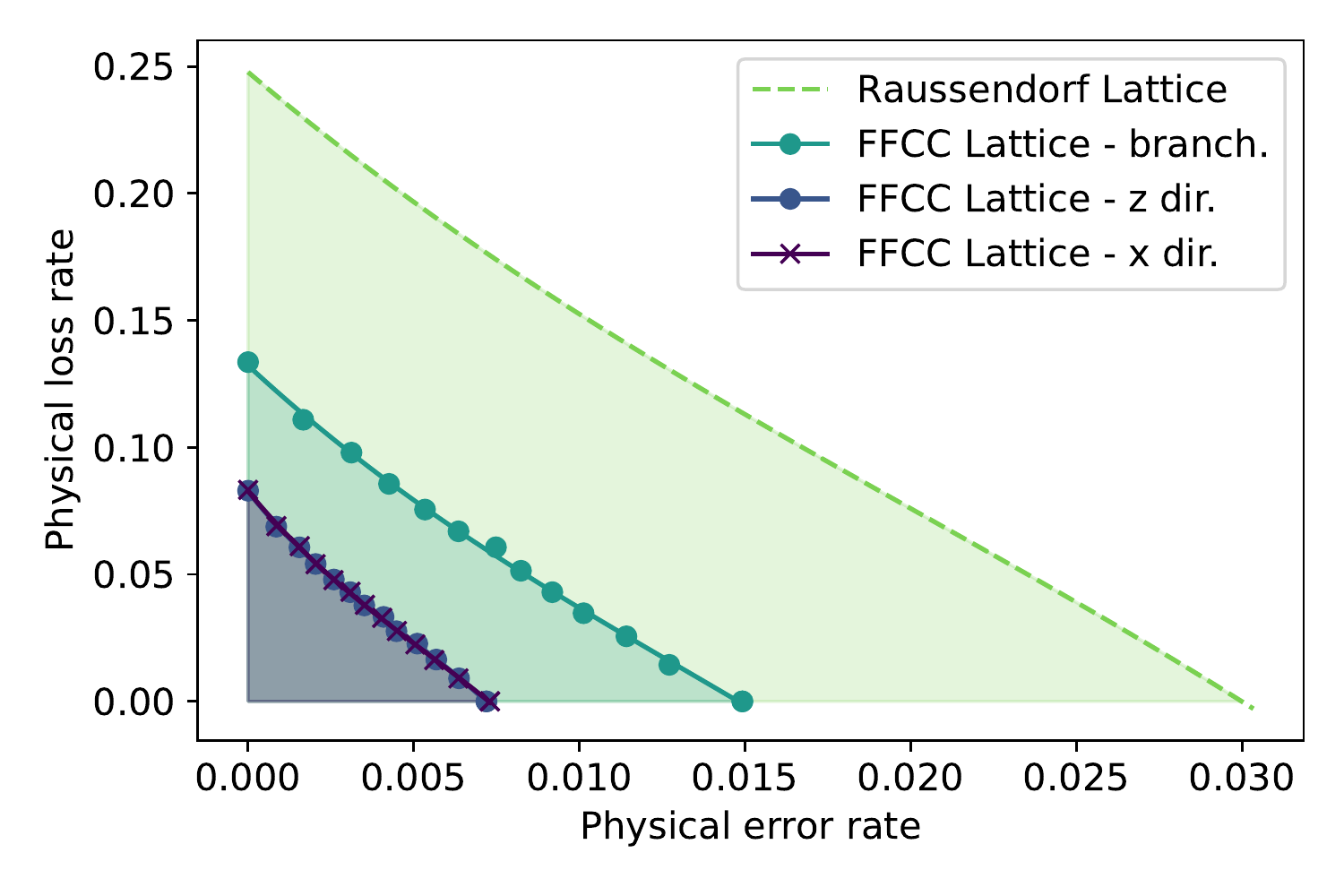}
  \caption{
Threshold error rates for the measurement based model under an unweighted independent and identically distributed noise model for both Pauli errors and loss. 
The correctable regions in the phase diagram are shown for the FFCC lattice for time-like (Z direction, dark blue) and space-like (X direction, purple with crossed marks) correlation surfaces shown in Fig.~\ref{fig:CorrelationSurfaces}, the branched FFCC lattice obtained transforming the FFCC via X-measurements on the (now virtual) two-valent nodes (dark green), and the topological cluster state model on the Raussendorf lattice~\cite{raussendorf2007} is shown by a dashed line. 
In all lattices, the thresholds for time-like and space-like correlation surfaces present identical behavior, and are thus shown only for the FFCC lattice, in which they in fact overlap.
}
  \label{fig:thresholds_iid}
\end{figure}

Here we present threshold estimates for the FFCC model with periodic boundary conditions with an independent and identically distributed noise model for both qubit erasure and Pauli bit-flip errors. We consider both a weighted- and an unweighted-noise model where error rates are weighted according to the valency of the underlying graph state~\cite{nickerson2018}.
We evaluate thresholds by measuring logical failure rates for varying system sizes and noise parameters.
For each combination of parameters we perform $10^4$ Monte-Carlo runs to estimate the logical error rate, see Appendix~\ref{app:decoding} for more details. 
In Fig.~\ref{fig:thresholds_iid} we present thresholds for models undergoing an unweighted independent and identically distributed noise model, delimiting the region in parameter space where the model is below threshold.
Considering bare lattices, the thresholds obtained with our model are lower than those of the topological cluster state model on the Raussendorf lattice~\cite{raussendorf2007, Barrett2010}, also reproduced in Fig.~\ref{fig:thresholds_iid}. 
These results are perhaps unsurprising, given that the FFCC has demonstrated a lower noise threshold compared to the surface code under a circuit-based noise model~\cite{ Kesselring2022}.
Thresholds are improved when considering the lattice for the FFCC being built via branched chains. 
This is because we effectively remove the two-valent nodes from the lattice. Nevertheless, measurement-based quantum computation still demonstrates lower thresholds than those of the topological cluster-state model on the Raussendorf lattice.
For completeness, we compare the logical error rate for different orientations of the correlation surface, with orientations that lie orthogonal to a canonical spatial direction, and the time-like direction.
We observe that thresholds are the same for  both types of correlation surface, see Figs.~\href{fig:CorrelationSurfaces}(b) and~(c).

We find then a discrepancy between the results obtained using the fusion-based noise model, and the independent and identically distributed noise model studied here.
Unlike the fusion-based noise model, the independent and identically distributed noise model does not account for noise that accumulates in the construction of the lattice~\cite{nickerson2018}. 
The fusion-based description of measurement-based quantum computing~\cite{Bartolucci2021fusion} provides a better model, as it facilitates a direct description of probabilistically fusing small resource states with linear optics, and enables a simple treatment of failed fusion operations and qubit loss and errors when building fault-tolerant structures.

\begin{figure}[]
  \centering
  \includegraphics[
  width=0.48 \textwidth]{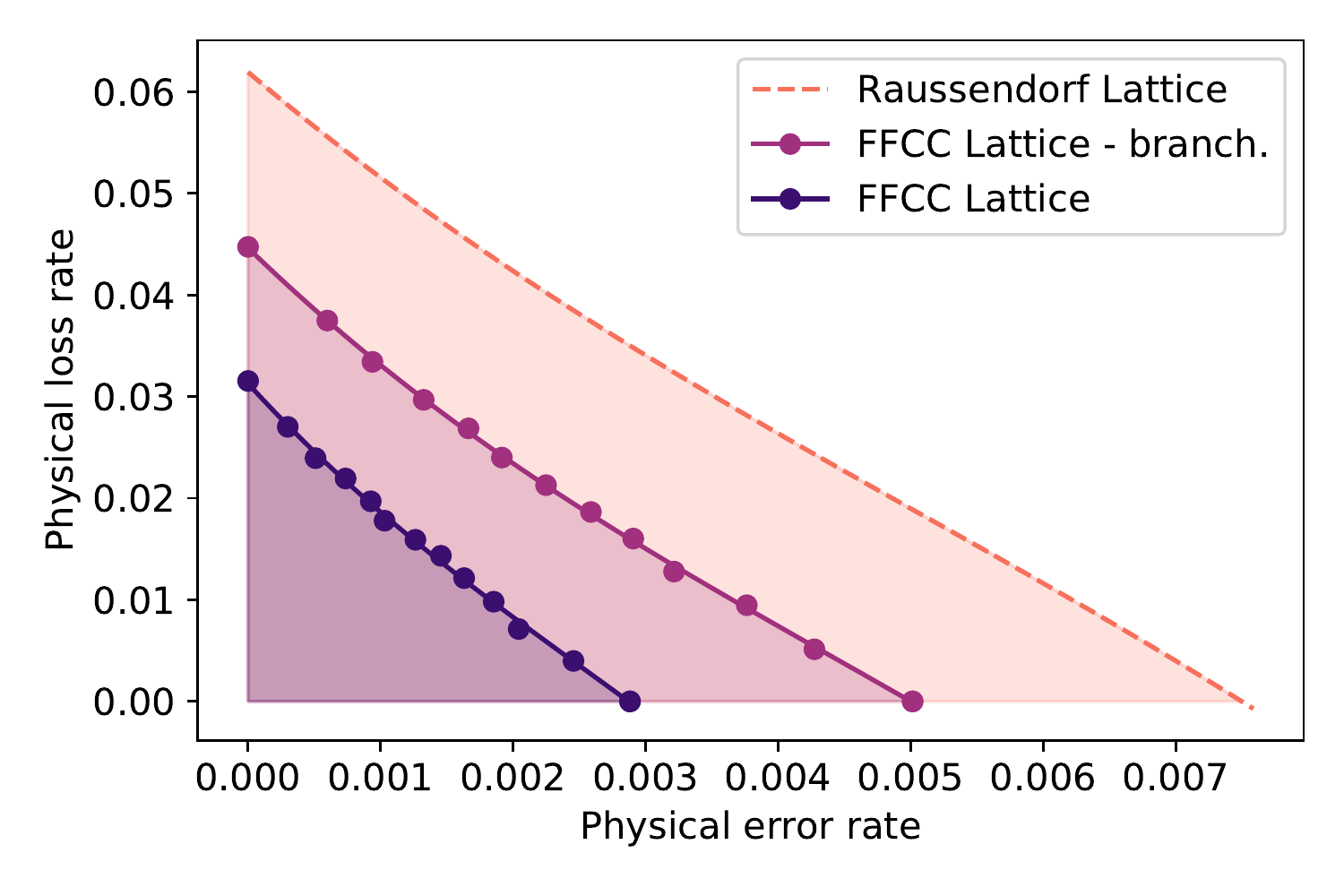}
  \caption{
Fault-tolerance performance of the bare lattices using the weighted Pauli error model in the simulations. 
The lattices and configurations considered are the same as in Fig.~\ref{fig:thresholds_iid}, with the considered error model being the only difference.}
  \label{fig:thresholds_wiid}
\end{figure}

Finally, we study the weighted-independent and identically distributed noise model.
This noise model partially accounts for the complexity of a lattice construction while still being platform-agnostic. Here, the error probability for each qubit in the lattice is multiplied by its valency~\cite{nickerson2018}.
The threshold estimates for the bare lattices using the weighted error model are shown in Fig.~\ref{fig:thresholds_wiid}.
The results obtained are qualitatively similar to those obtained in Fig.~\ref{fig:thresholds_iid}, where now the difference in performance between the FFCC lattice and the topological cluster state model on the Raussendorf lattice is smaller compared with the study using the unweighted phenomenological error model.
This is another indication that, because the FFCC lattice has been designed for ease of practical construction rather than of the bare lattice itself, the closer the error model gets to practical implementations the better the performance of FFCC becomes, relative to other well-studied constructions.

\section{Simulation methods}
\label{app:decoding}

\begin{figure*}[]
  \centering
  \includegraphics[
  width=1.0 \textwidth]{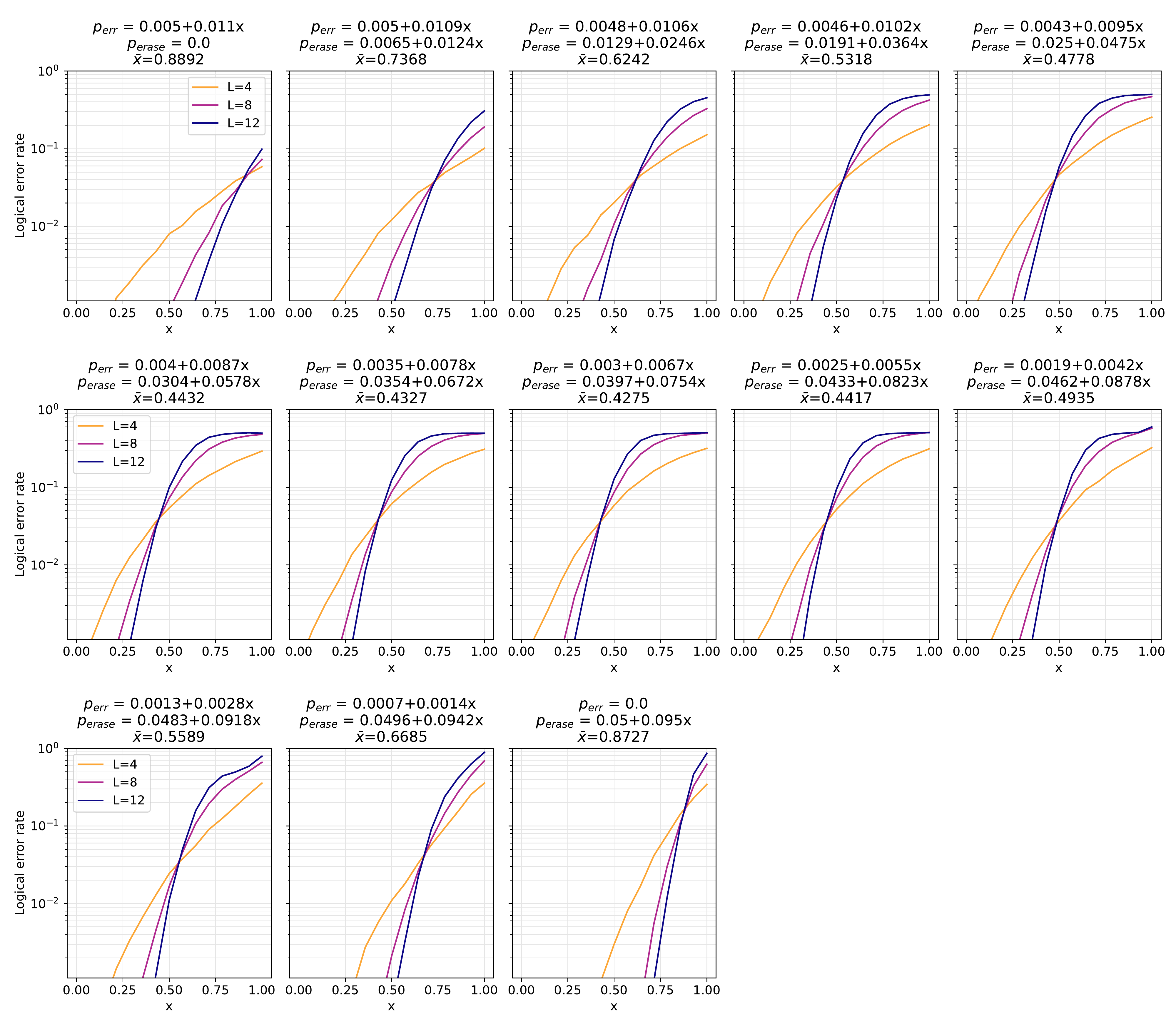}
  \caption{
Examples of threshold estimates along linear paths $(p_\text{erase}, p_\text{err}) = (x p^0_\text{erase}, x p^0_\text{err})$ in the two-dimensional erasure-error phase space. The data reported here are for the fusion-based construction of the FFCC lattice using branched chains ($\ell \rightarrow \infty$ case), with each threshold representing a marker in the associated plot in Fig.~\ref{fig:thresholds_fusion}.}
  \label{fig:all_thresholds}
\end{figure*}

Here we provide details on the numerical implementation of noise models and decoders to estimate thresholds in both the simulations for measurement-based lattices and fusion-based constructions.
The source code used is openly accessible on GitHub~\cite{github_repo}.  
The simulations and decoding can be performed in the same way when considering both bare lattices and fusion networks.
In the first case, the decoding is performed using a matching matrix that describes the lattice qubits that contribute to each detector, while in the second case it instead describes the fusions that contribute to each detector in an analogous fashion.
The phenomenological independent and identicallly distributed error models also have identical descriptions via these matching matrices: erasure and error for a fusion outcome has the same effect of erasure and error for a qubit.
This means that simulating noise can be done in the same manner for both cases, it is only the interpretation of the results that differs.
To estimate the thresholds we perform Monte-Carlo simulations where, for given values of noise parameters, in each sample we model qubit erasures and errors for each single-qubit measurement (fusion outcomes) in the primal lattice (fusion network) independently.
Decoding is performed via a minimum-weight perfect matching(MWPM) decoder, which we implement with \texttt{PyMatching}~\cite{higgott2022}. 
The MWPM decoder provides a correction, which we combine with the real error to determine if a logical error has occurred. 
For each combination of noise parameters, we perform $10^4$ Monte-Carlo samples to estimate the logical error rate, and repeat it for various lattice sizes $L\in \{4, 8, 12\}$ to identify thresholds.
As the dual lattices and networks have an identical structure to the primal lattice for all the models we study, we focus on the primal lattice as the dual lattice will have the same performance. 
To calculate thresholds in the two-dimensional phase space with erasure and loss and identify the correctable regions, we perform linear scans of erasure and error values, obtained setting $(p_\text{erase}, p_\text{err}) = (x p^0_\text{erase}, x p^0_\text{err})$ and varying $x$.
For each value of $p^0_\text{erase}$ and $p^0_\text{err}$, a threshold is observed along the linear scan in the two-dimensional space, which corresponds to a single mark plotted in Figs.~\ref{fig:thresholds_iid},~\ref{fig:thresholds_fusion} and~\ref{fig:thresholds_wiid}.
Repeating this procedure for various values of $p^0_\text{erase}$ and $p^0_\text{err}$ allows us to reconstruct the fault-tolerant regions.
An example of this procedure for the fusion-based construction of the FFCC lattice using branched chains ($\ell \rightarrow \infty$ case) is shown in Fig.~\ref{fig:all_thresholds}.

\section{Improving thresholds}
\label{app:biased}

\begin{figure}[]
  \centering
  \includegraphics[
  width=0.45 \textwidth]{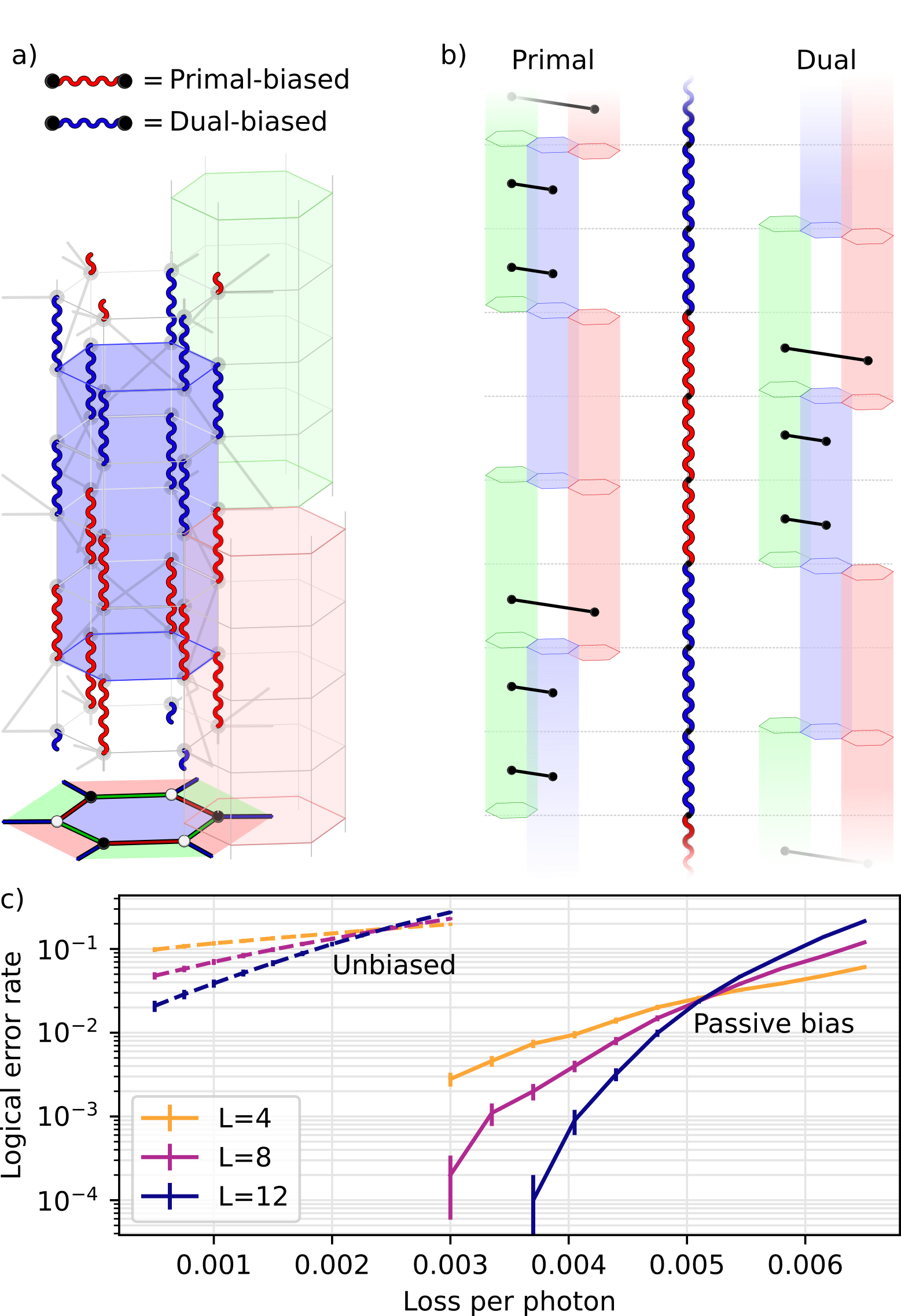}
  \caption{
(a) Passive fusion biasing configuration for the FFCC fusion-based construction with branched chains and $\ell=\infty$. Primal-biased (in red) fusions represent fusions configured to erase the fusion outcome associated with dual detectors upon failure but not the one used in primal detectors, and vice-versa  for dual-biased fusions (in blue).  
(b) Schematic of fusion failures connecting detectors in the primal and dual syndrome graphs considering the passive fusion biasing configuration as in (a). In both cases, no connections are present between red and blue detectors, implying erased fusion outcomes are restricted to disconnected two-dimensional plains if no loss is present.
(c) Loss threshold estimates for passively biased fusions (solid lines) and unbiased fusions (dashed lines), considering boosted fusions with $25\%$ failure rate and a loss-only error model.
}

  \label{fig:biasedfusions}
\end{figure}

We here discuss possible techniques that can be used to improve the fault-tolerance performance of the fusion-based constructions we present in the main text. 
For simplicity, and for consistency when comparing to previous fusion-based models from Ref.~\cite{Bartolucci2021fusion}, in the main text we considered an unbiased fusion failure model, where each of the parity outcomes $XX$ and $ZZ$ from a fusion is erased with the same $p_\text{erase}(XX) = p_\text{erase}(ZZ) = 1 - (1- p_\text{fail}/2)  \eta^{1/p_\text{fail}}$, where $p_\text{fail}$ is the fusion failure probability, $\gamma$ is the total optical loss, and $\eta = 1 - \gamma$ is the transmission.  
However, more in general, such failure probability between the two parity measurements can be readily biased by physically adjusting the reflectivity of two beam-splitters, one for each fused photon, in the linear optical circuits implementing fusion, with unbiased noise corresponding to randomly choosing between reflectivities $(0, 0)$ and $(\frac{1}{2}, \frac{1}{2})$ ~\cite{Bartolucci2021fusion, sahay2022tailoring, bombin2023increasing}.
Very recently, it was shown that this capability to program the bias in fusion failures can bring significant performance enhancements compared to the unbiased cased. 
In particular, Ref.~\cite{sahay2022tailoring} showed that for  Raussendorf lattice fusion constructions a passive bias where fusion failure is entirely weighted on just one of the two parity outcomes, i.e. outcome-dependent erasure probabilities  $p_\text{erase}(XX) = 1 - (1- p_\text{fail})  \eta^{1/p_\text{fail}}$ and $p_\text{erase}(ZZ) = 1 -   \eta^{1/p_\text{fail}}$ or vice-versa, can significantly improve the fault-tolerant threshold relative to photon loss.
Such advantages arise from mechanisms analogous to the XZZX surface code constructions~\cite{bonilla2021xzzx, Darmawan2021practical, Claes2022}: the idea is to exploit the bias to restrict errors in 2-dimensional cuts of the 3-dimensional code, where they can be decoded with higher success probability.  
Furthermore, Ref.~\cite{bombin2023increasing} also showed that further improvements, by an additional factor $\sim 3$ can be obtained by adding adaptivity in the failure bias choice.
These advantages come without any additional hardware overhead compared to the unbiased case, although the adaptive case requires fast classical processing of the fusion outcomes, feedback, and reprogramming of the fusion circuits.

In Refs.~\cite{sahay2022tailoring, bombin2023increasing}
the biasing techniques were demonstrated for the fusion constructions based on the Raussendorf lattice from Ref.~\cite{Bartolucci2021fusion}.
These techniques are readily adapted to our construction as well, as we show here.
Analogously to Ref.~\cite{sahay2022tailoring}, we consider passive biasing of fusion failures and show how it can readily be implemented with our model to improve loss thresholds.
In Fig.~\hyperref[fig:biasedfusions]{\ref{fig:biasedfusions}a} the configuration we use to bias the fusions is described, with an alternation of three layers with failure fully biased on the primal fusion outcomes (i.e. fusion failure always erases only the fusion outcome associated to the dual detectors) and three layers with failure fully biased on the dual fusion outcomes.
In this way, in the low loss regime, both in the primal and dual lattices, erasure errors due to fusion failures are not transmitted between the blue and red detection cells (see Fig.~\hyperref[fig:biasedfusions]{\ref{fig:biasedfusions}b}), resulting in the restriction of fusion erasures within two-dimensional layers.

In Fig.~\hyperref[fig:biasedfusions]{\ref{fig:biasedfusions}c} we report the fault-tolerance thresholds relative to photon loss with biased and unbiased fusions for our construction with branched chains.
Here, we consider a loss-only noise model, in analogy with Ref.~\cite{sahay2022tailoring}, and a physical boosted fusion gate with failure probability $p_\text{fail}=25\%$~\cite{Grice2011}.
A significant improvement, by more than a factor of two, is obtained for the loss threshold for the passive-bias case ($0.52\%$) compared to no bias ($0.24\%$). 
For the same noise model and physical fusions with failure probability $p_\text{fail}=25\%$, Ref.~\cite{sahay2022tailoring} obtains a threshold for the fusion-based construction of the Raussendorf lattice from  six-ring resource states can tolerate up to $0.37\%$, indicating the FFCC construction provides improved thresholds also in when biased fusions are utilized. 
These performance enhancements are summarised in Table~\ref{tab:summary}.
Adaptive approaches such as those investigated in Refs.~\cite{Auger2018, bombin2023increasing} can then also be explored to further improve the loss threshold for our constructions, which we will investigate in future works.

\section{Noise sources in resource state generation}
\label{app:physical_noise}

\begin{figure*}[]
  \centering
  \includegraphics[
  width=0.98 \textwidth]{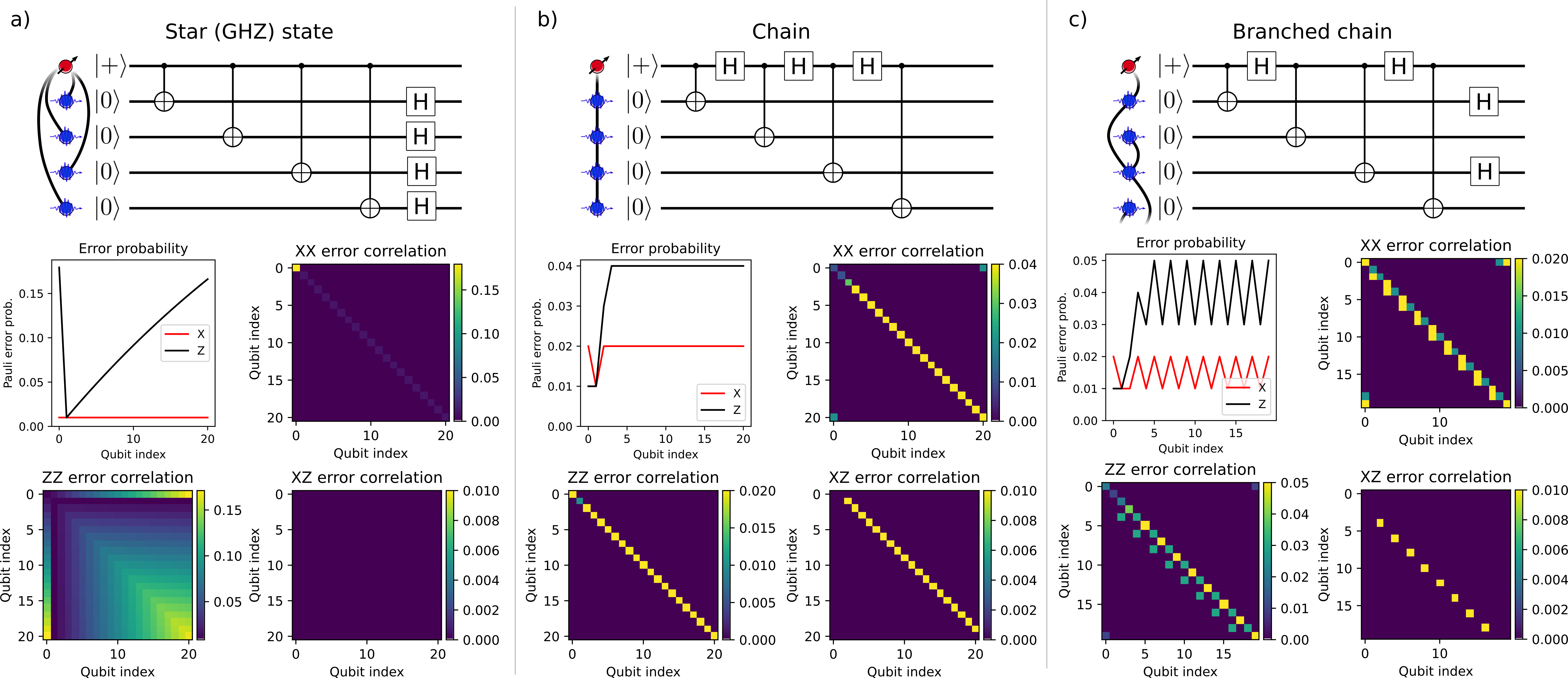}
  \caption{
Circuits associated with the graph state generation for star graphs (a) locally equivalent to GHZ states, chains (b), and branched chains (c) are shown, together with Pauli $X$ and $Z$ error probabilities for individual qubits and error correlation matrices between different qubits.
In all plots, the spin qubit (shown in red in the circuits) has qubit index zero, while photonic qubits (shown in blue in the circuits) have indexes which follow their order of generation.
All plots consider resource states a single spin qubit and 20 photonic qubits.
}
  \label{fig:QDnoises}
\end{figure*}

Errors in the outcomes of fusion measurements can arise from a variety of physical processes, e.g. imperfect resource state preparation, photon distinguishability, etc. 
In the main text, for simplicity, we used a phenomenological model where all such noises are taken into account into a single parameter, the ``fusion error rate'', and fusion errors are assumed to be independent and identically distributed distributed.
In general, physical errors will differ from such a model to some degree, e.g. by inserting correlations between separate fusions.
We here qualitatively discuss errors in the generation of the resource states with quantum emitters to argue that, when considering more detailed error models, one-dimensional cluster states are preferable compared to other types of resource states generatable with quantum emitters due to maintaining locality in the generated correlations. 
This observation was already presented for chains by Linder and Rudolph in Ref.~\cite{Lindner2009proposal}, with a detailed error model for a quantum emitter based on a quantum dot.
Here we extend a similar investigation to branched chains and GHZ-type (star graphs) states, which represent all possible classes of resource graph states (up to local transformations) generatable with a single quantum emitter~\cite{Li2022}. 
Fig.~\ref{fig:QDnoises} shows the circuits that correspond to the generation of GHZ states, chains, and branched chains with a spin-photon interface in a quantum emitter Ref.~\cite{Lindner2009proposal}. 
The spin-dependent photon generation corresponds to CNOT operations between the spin qubit and the photonic qubits initially prepared in the $\ket{0}$ state, and different types of states are obtained applying Hadarmad gates on the spin qubit (corresponding to $\pi/2$ pulses) between consecutive photon generations.
To have a qualitative description, we simulated errors arising in the resource state generation by considering a circuit noise model for such circuits where $T_1$ and $T_2$ noise processes of the spin are simulated via a bit- and phase-flip channel adding an $X$ and a $Z$ Pauli error to the spin qubit between the generation of two consecutive photons independently and with probability $p_1$ and $p_2$, respectively. 
Noisy gate operations on the spin are also simulated by applying a perfect operation followed by independent bit- and phase-flip channels, both adding a Pauli error with probability $p_{gate}$.
For simplicity, we considered $p_1 = p_2 = p_{gate} = p$ to describe the system with a single noise parameter $p$.
In Fig.~\ref{fig:QDnoises} we report the final probability to have a Pauli $X$ and $Z$ error on each qubit (where the spin is always labeled with index zero), as well as the correlation matrices for Pauli errors between qubits, obtained with the noise model described above. 
It can be observed that, while for GHZ-type states very non-local error correlations are present, for the one-dimensional resource states (chains and branched chains) correlations are always restricted between neighboring or next-neighboring photons, consistently with the observations in Ref.~\cite{Lindner2009proposal}.
This indicates that if fusions are applied locally, as in our constructions, such locality will be maintained when using chains and branched chains as resource states, so errors will not spread rapidly in the lattice.
This is not necessarily the case for GHZ-states due to the arising non-local correlations, suggesting further advantages for one-dimensional resource states when considering more physical noise models. 
A full analysis of fault-tolerance performance for physical noise models specific to hardware platforms for quantum emitters will be investigated in future work.

\begin{table*}[b]
  \centering
\begin{tabular}{|c||c|c||c|c|}
\hline
\thead{Fusion Lattice} & \thead{Threshold:\\fusion errors} & \thead{Threshold:\\fusion erasure} & \thead{Threshold:\\photon loss\\(unbiased, $p_\text{fail}= 25\%$)}  & \thead{Threshold:\\photon loss\\(biased, $p_\text{fail}= 25\%$)}\\
\hhline{|=||=|=||=|=|}
Raussendorf - stars~\cite{Bartolucci2021fusion} & 0.7\% & 6.3\% & - & -\\
\hline
Raussendorf - hexagons~\cite{Bartolucci2021fusion,sahay2022tailoring} & 1.0\% & 11.9\% & - & 0.37\% \\
\hline
FFCC - branched chains & 1.5\% & 13.3\% & 0.24\% & 0.52\% \\
\hline
\end{tabular}
  \caption{
    Summary of performances of different fusion-based fault-tolerant constructions in terms of thresholds.
  }
  \label{tab:summary}
\end{table*}

\end{document}